# Site preference of transition-metal elements in $L1_0$-TiAl: A first-principles study

Chao Jiang[*]

Structure/Property Relations Group (MST-8), Los Alamos National Laboratory, Los Alamos, New Mexico 87545

## ABSTRACT

The site preference of $3d$ (Ti-Cu), $4d$ (Zr-Ag) and $5d$ (Hf-Au) transition-metal elements in $L1_0$ TiAl is studied using a combination of first-principles supercell calculations and the statistical mechanical Wagner-Schottky model. Our key finding is that both temperature and alloy stoichiometry can strongly affect the site occupancy behavior of ternary alloying elements in $L1_0$ TiAl. We further predict that the tendency of transition metals to occupy the Al sites in TiAl increases with increasing $d$-electron number along a series.

*Corresponding author: chao@lanl.gov

TiAl intermetallic compound has been considered a promising high-temperature structural material due to its low density, adequate high temperature strength, and good oxidation resistance [1]. Its practical application is nevertheless hindered by the lack of room temperature ductility, presumably due to the covalent nature of Ti-Al bonding [2]. One important way to enhance the ductility of TiAl is through alloying with ternary elements [1, 2]. In understanding the role of those alloying elements in modifying the mechanical properties of TiAl, knowledge of their site occupancy behavior is indispensable. The present work aims at predicting the site preference of $3d$ (Ti-Cu), $4d$ (Zr-Ag) and $5d$ (Hf-Au) transition-metal elements in TiAl using a combination of statistical mechanics and first-principles calculations based on density functional theory. The effects of both alloy stoichiometry and temperature on site preference will be fully considered. As will be shown, our calculated results compare favorably with existing experimental and theoretical studies in the literature.

TiAl has an ordered $L1_0$ (AuCu-type) structure in which Ti and Al atoms alternatively occupy the (002) planes of a tetragonally distorted fcc unit cell. Deviations from the ideal AB stoichiometry are accommodated by the creation of structural (constitutional) point defects, *e.g.*, antisites ($Ti_{Al}$, $Al_{Ti}$) or vacancies ($Va_{Al}$, $Va_{Ti}$). Here $i_\alpha$ denotes species $i$ on the α sublattice and Va denotes vacancy. In a ternary $L1_0$ TiAl-X alloy, two new types of point defects are introduced since ternary element X can occupy either Ti ($X_{Ti}$) or Al ($X_{Al}$) sites. In our study, it is assumed that the defect concentrations are sufficiently dilute to allow use of the Wagner-Schottky model [3]. We adopt in our treatment a canonical ensemble [4] containing a fixed number of atoms. In considering the finite temperature effects, the vibrational entropy is neglected due to limited computing resources and only mean-field configurational entropy is considered. A detailed description of the general formalism for treating site preference of ternary elements in ordered compounds with AB stoichiometry has already been given previously [5, 6] and will not be repeated here.

A first-principles supercell approach [4-10] is employed to obtain the formation energies of isolated point defects in $L1_0$-TiAl. We use 32-atom 2×2×2 supercells, each containing a single point defect (vacancy, antisite or ternary element) at its center. First-principles calculations are performed using projector augmented wave [11] pseudopotentials within the generalized gradient approximation (PW91-GGA) [12], as implemented in Vienna *ab initio* simulation package (VASP) [13]. The semi-core $3p$ electrons of Ti, V, Cr and Mn, the semi-core $4p$ electrons of Nb, Mo and Tc, the semi-core $5p$ electrons of Hf, Ta and W, and both the semi-core $4s$ and $4p$ electrons of Zr are explicitly treated as valence. The plane wave cutoff energy is set at 350 eV. The *k*-point meshes for Brillouin zone sampling are constructed using the Monkhorst–Pack scheme [14]. For the 32-atom supercells, a 9×9×9 *k*-point mesh corresponding to 75 irreducible *k*-points in the Brillouin zone is sufficient to give fully converged results. Spin-polarized calculations are performed for Cr, Mn, Fe, Co, and Ni. According to our calculations, Co, Fe and Mn develop large local magnetic moments (>1.5$\mu_B$) when they occupy the Ti sites in TiAl. Using a conjugate-gradient scheme, each supercell is fully relaxed with respect to unit cell volume, shape (*c/a* ratio), as well as all internal atomic positions. Our predicted equilibrium lattice constants of the ideal $L1_0$-TiAl structure (*a*=3.987 Å, *c*=4.072 Å, *c/a*=1.02) are in excellent agreement with the experimental values of $a$=3.99 Å and $c$=4.07Å [15]. Finally, we obtain the defect formation energies using finite differencing [5, 6, 16].

Before addressing the site preference behavior of ternary alloying elements in TiAl, it is necessary to first have a good understanding of the types of intrinsic point defects in binary TiAl. At T=0K, the point defect structure of an ordered compound is solely governed by enthalpy and the point defects stable at this temperature are called structural defects. In accordance with previous experimental [17, 18] and theoretical [19, 20] studies, Table 1 shows that antisite defects have much lower formation energies than vacancies in TiAl, thus indicating that the structural defects in TiAl are antisites on both sides of stoichiometry. The lower formation energy of Al antisites than that of Ti antisites may also explain why the homogeneity range of TiAl is significantly shifted toward Al-rich compositions in the equilibrium Ti-Al phase diagram [21].

Next, we consider the T=0K site occupancy behavior of a ternary element X in TiAl. For a Al-rich $Ti_{0.5-x}Al_{0.5}X_x$ alloy, there are two possible lattice configurations depending on the location of X atoms: (i) X occupies the Ti sites and the configuration is simply (Ti,X)(Al). (ii) X occupies the Al sites at the cost of forming Al antisites to yield the configuration (Ti,Al)(Al,X). The creation of vacancies is not considered since they are energetically very unfavorable compared to antisites. It is the energy difference between those two configurations, $E_i - E_{ii} = H_{X_{Ti}} - H_{X_{Al}} - H_{Al_{Ti}}$, that determines the site preference of X in TiAl. Here $H_d$ is the formation energy of an isolated point defect of type $d$. X preferentially occupy the Ti sites when $E_i < E_{ii}$ and vice versa. Similarly, there are also two possible lattice configurations for a Ti-rich $Ti_{0.5}Al_{0.5-x}X_x$ alloy: (i) X occupies the Ti sites at the cost of forming Ti antisites to yield the configuration (Ti,X)(Al,Ti), (ii) X occupies the Al sites and the configuration is simply (Ti)(Al,X). The energy difference between the two configurations now becomes $E_i - E_{ii} = H_{X_{Ti}} - H_{X_{Al}} + H_{Ti_{Al}}$. Finally, for a stoichiometric $Ti_{0.5-x/2}Al_{0.5-x/2}X_x$ alloy, there are three possible lattice configurations: (i) X occupies the Ti sublattice accompanied by the formation of Ti antisites and the configuration is (Ti,X)(Al,Ti). (ii) X occupies the Al sublattice accompanied by the creation of Al antisites and the configuration is (Ti,Al)(Al,X). (iii) X randomly occupies both Al and Ti sublattices to yield the (Ti,X)(Al,X) configuration, in which 1/2 of the X atoms occupy the Ti sites and the remaining 1/2 occupy the Al sites. The energy differences between configurations (i) and (iii) and between configurations (iii) and (ii) are $E_i - E_{iii} = (H_{X_{Ti}} - H_{X_{Al}} + H_{Ti_{Al}})/2$ and $E_{iii} - E_{ii} = (H_{X_{Ti}} - H_{X_{Al}} - H_{Al_{Ti}})/2$.

Based on the above analysis, the T=0K site occupancy behavior of substitutional ternary elements in TiAl can be classified into the following three general types:

(i) $H_{X_{Ti}} - H_{X_{Al}} + H_{Ti_{Al}} < 0$: X always prefers the Ti sites in TiAl independent of composition, *i.e.*, X exhibits strong Ti site preference.

(ii) $H_{X_{Ti}} - H_{X_{Al}} - H_{Al_{Ti}} > 0$ (or equivalently $H_{X_{Ti}} - H_{X_{Al}} + H_{Ti_{Al}} > H_{Ti_{Al}} + H_{Al_{Ti}}$): X always prefers the Al sites in TiAl independent of composition, *i.e.*, X exhibits strong Al site preference.

(iii) $0 < H_{X_{Ti}} - H_{X_{Al}} + H_{Ti_{Al}} < H_{Ti_{Al}} + H_{Al_{Ti}}$: The site preference of X is strongly composition-dependent. X prefers the Ti sites in Al-rich TiAl and the Al sites in Ti-rich TiAl, and shows no site preference in stoichiometric TiAl by randomly occupying both Al and Ti sites.

It is worth noting that a similar methodology has been employed by Ruban and Skriver [22] in classifying the site substitution behavior of ternary additions to Ni$_3$Al. Clearly, the T=0K site preference of any substitutional ternary element in TiAl can be completely characterized by a single parameter $E_X^{Al \to Ti} = H_{X_{Ti}} - H_{X_{Al}} + H_{Ti_{Al}}$, which has the physical meaning as the energy required in transferring a X atom from a Al site to a Ti site via the reaction: $X_{Al} + Ti_{Ti} \to X_{Ti} + Ti_{Al}$. Its value can be directly obtained from first-principles calculated total energies of four 32-atom supercells as follows:

$$E_X^{Al \to Ti} = E(Ti_{15}Al_{16}X) - E(Ti_{16}Al_{15}X) + E(Ti_{17}Al_{15}) - E(Ti_{16}Al_{16}) \qquad (1)$$

Our calculated values of $E_X^{Al \to Ti}$ for each of the 3$d$ (Ti-Cu), 4$d$ (Zr-Ag) and 5$d$ (Hf-Au) transition-metal elements are summarized in Fig. 1. The formation energy of the exchange antisite defect in TiAl, $H_{Ti_{Al}} + H_{Al_{Ti}}$, is calculated to be 1.01 eV. According to Fig. 1, Zr and Hf fall into type (i), Co, Ru, Rh, Pd, Ag, Re, Os, Ir, Pt and Au fall into type (ii), and V, Cr, Mn, Fe, Ni, Cu, Nb, Mo, Tc, Ta and W fall into type (iii). The general trend is that the preference for the Al sublattice increases with increasing $d$-electron number along a series. Such a trend is obeyed by all 3*d*, 4*d* and 5*d* elements except for those at the end of a series.

It should be noted that the above classification is strictly valid only at T=0K. At finite temperatures, entropy will also play an important role in determining the site preference of ternary elements. As the consequence, site preference reversal may occur with increasing temperature. In Fig. 2, the fraction of X atoms occupying the Al sites in TiAl alloys containing 1% of X at 1173K is plotted as a function of Ti concentration. Fig. 3 further shows the equilibrium partitioning of X atoms between Al and Ti sites in Al-rich, Ti-rich and stoichiometric TiAl alloys as a function of temperature. The horizontal dashed lines indicate random occupation of Al and Ti sites by X atoms, *i.e.*, no site preference. At all alloy compositions and temperatures, type (i) and type (ii) elements consistently show a predominant preference for the Ti and Al sublattices, respectively. In contrast, the site preference of type (iii) elements exhibits a complex dependence on both composition and temperature.

The site occupancy of Nb, Mo, Ta and W in TiAl at T=1473K has been predicted by Woodward *et al.* [20] using first-principles methods. Their study showed that the site preference of Mo and W changes with alloy stoichiometry, which is in agreement with our classification of Mo and W as type (iii) elements. Interestingly, although Nb and Ta are both type (iii) elements and thus exhibit strongly composition-dependent site preference at T=0K, they actually show a consistent preference for the Ti sublattice over the whole composition range at high temperatures, although the preference is weaker than that of type (i) elements (see Fig. 2). Again, such a conclusion agrees with the study by Woodward *et al*. [20].

Using X-ray diffraction and atom probe techniques, Kim *et al*. [23] observed that Ru strongly prefers the Al sites in TiAl. Their results are in support of our classification of Ru as a type (ii) element. Using atom location by channeling enhanced microanalysis (ALCHEMI), Chen *et al*. [24, 25] and Rossouw *et al*. [26] found that Zr and Hf exclusively occupy the Ti sites, while Cr, Mn and Mo occupy both Al and Ti sites in TiAl. Those experimental evidences are consistent with our classification of Zr and Hf as type (i) elements and of Cr, Mn and Mo as type (iii) elements. Using the same technique, Hao *et al*. [27] observed that the site preference of V, Cr and Mn changes considerably with alloy composition, which can be understood since V, Cr and Mn are all type (iii) elements. They also found that Nb consistently prefers the Ti sites while Fe consistently prefers the Al sites regardless of alloy composition. Although Fe and Nb are both type (iii) elements, the experimental observations can be explained by the fact that Fe and Nb actually behave like a type (ii) and type (i) element at high temperatures, respectively (see Fig. 2). Finally, Fig. 4 shows our predicted site occupancy of Ta, Zr, Nb, Mo, V, Cr, Mn, Fe and Ni in $Ti_{0.47}Al_{0.51}X_{0.02}$ alloys in direct comparison with the experimental data from Hao *et al*. [27]. All calculations are performed at the experimental annealing temperature of 1173K. With the exception of Cr and Fe, the agreement is quite satisfactory. The origin of such discrepancies may be due to large experimental uncertainties as well as the neglect of vibrational entropy in our calculations.

In summary, using a combination of first-principles calculations and a statistical-mechanical Wagner-Schottky model, we predict the site preference of *3d*, *4d* and *5d* transition-metal elements in L1$_0$ TiAl as a function of both alloy stoichiometry and temperature. At all alloy compositions and temperatures, Zr and Hf have a predominant preference for the Ti sites, while Co, Ru, Rh, Pd, Ag, Re, Os, Ir, Pt and Au have a predominant preference for the Al sites. For V, Cr, Mn, Fe, Ni, Cu, Nb, Mo, Tc, Ta and W, their site preference will depend quite sensitively on alloy stoichiometry as well as heat treatment, *e.g.*, annealing temperature and cooling rate.

## ACKNOWLEDGEMENTS

This work is financially supported by Director's postdoctoral fellowship at Los Alamos National Laboratory (LANL). All calculations were performed using the parallel computing facilities at LANL.

Table 1. First-principles calculated formation energies (eV/defect) of the intrinsic point defects in L1$_0$ TiAl. Reference states: fcc Al and hcp Ti.

| Defect | Ti$_{Al}$ | Al$_{Ti}$ | Va$_{Al}$ | Va$_{Ti}$ |
|---|---|---|---|---|
| Formation Energy | 0.84 | 0.17 | 2.20 | 1.46 |

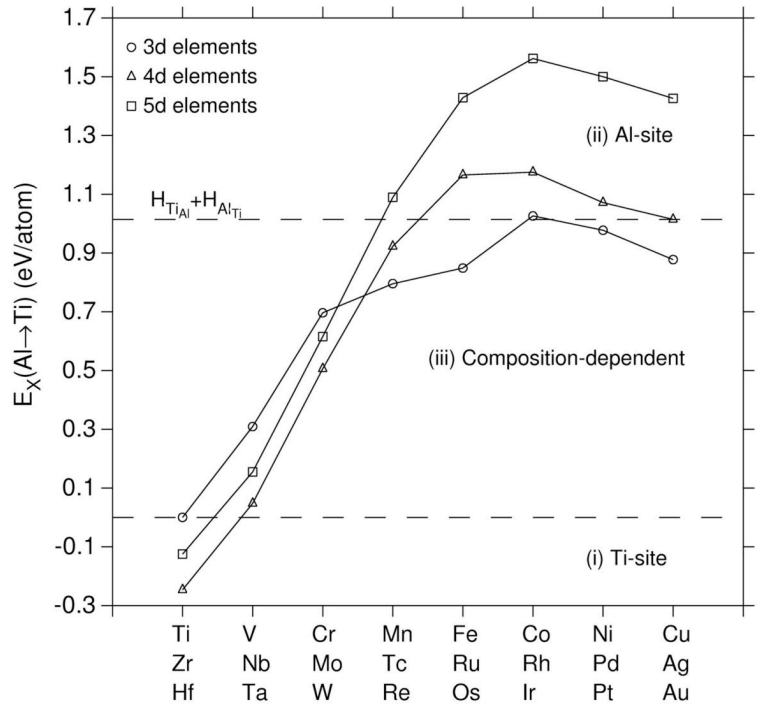

Fig. 1. Classification of the T=0K site preference behavior of 3$d$, 4$d$ and 5$d$ transition-metal elements in L1$_0$ TiAl based on $E_X^{Al \to Ti}$ obtained from first-principles calculations on 32-atom supercells.

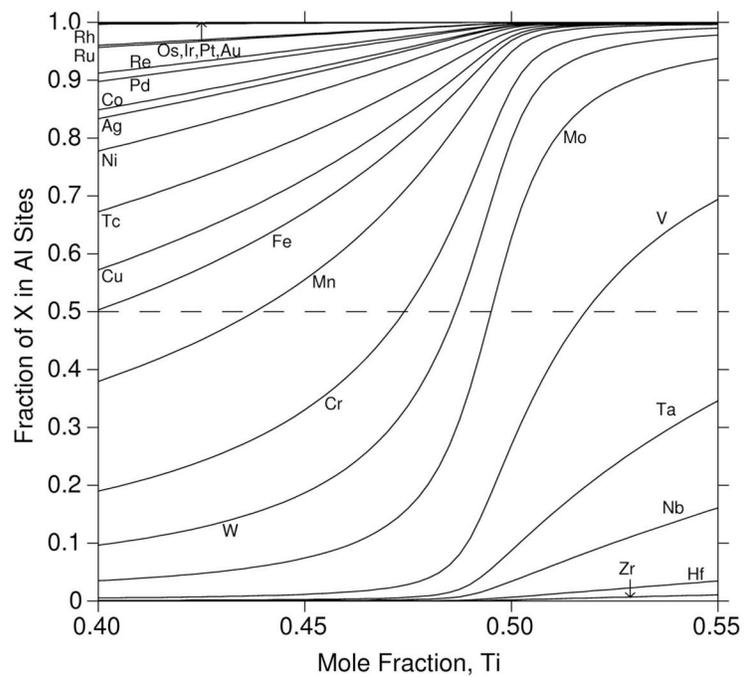

Fig. 2. Predicted fraction of X occupying the Al sublattice in Ti$_x$Al$_{0.99-x}$X$_{0.01}$ alloys at T=1173K as a function of Ti concentration.

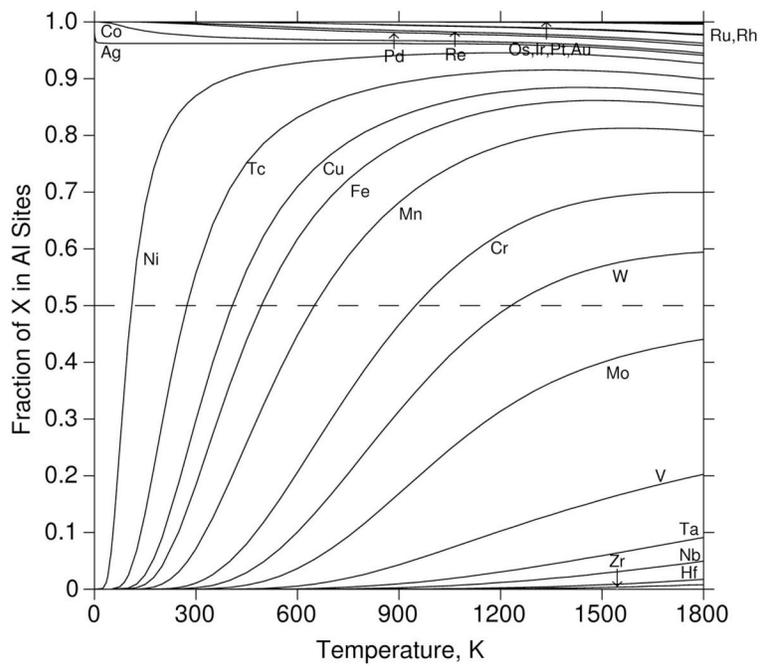

(a)

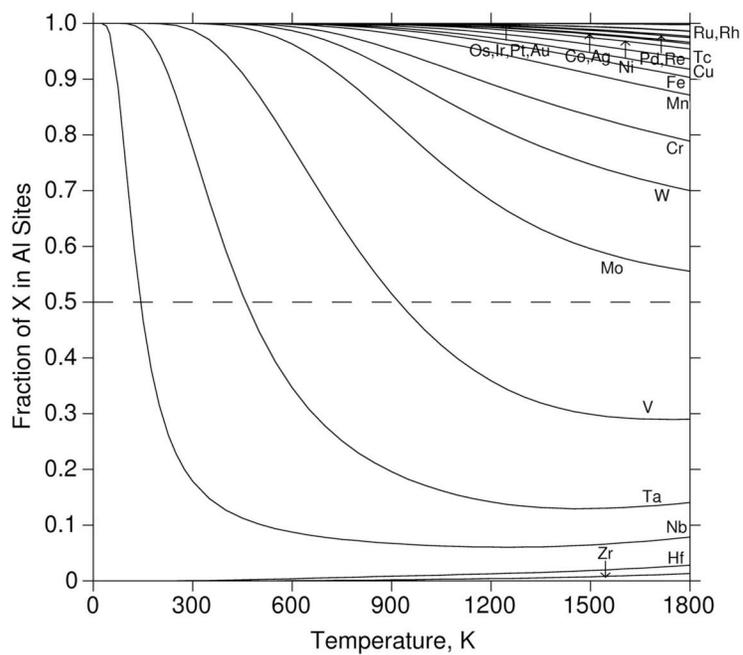

(b)

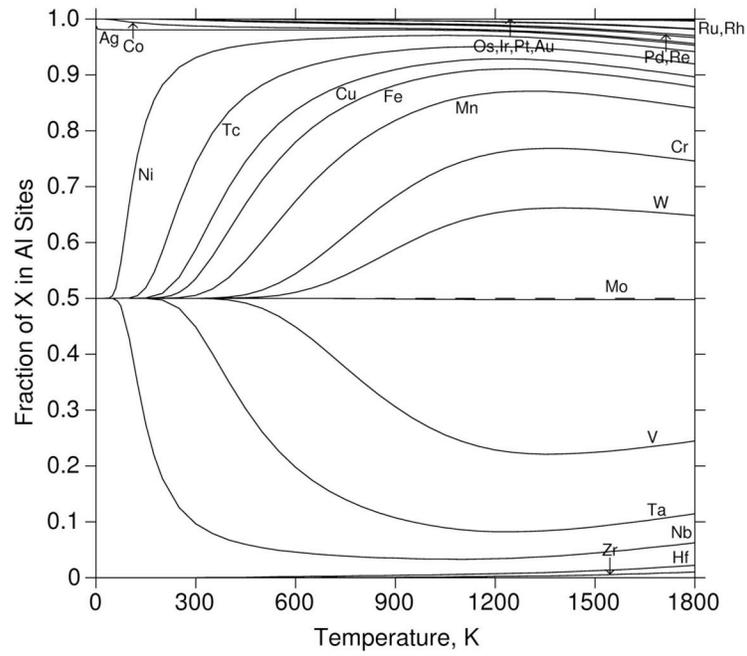

(c)

Fig. 3. Predicted fraction of X occupying the Al sublattice in (a) Al-rich $Ti_{0.48}Al_{0.5}X_{0.02}$, (b) Ti-rich $Ti_{0.5}Al_{0.48}X_{0.02}$, and (c) stoichiometric $Ti_{0.49}Al_{0.49}X_{0.02}$ alloys as a function of temperature.

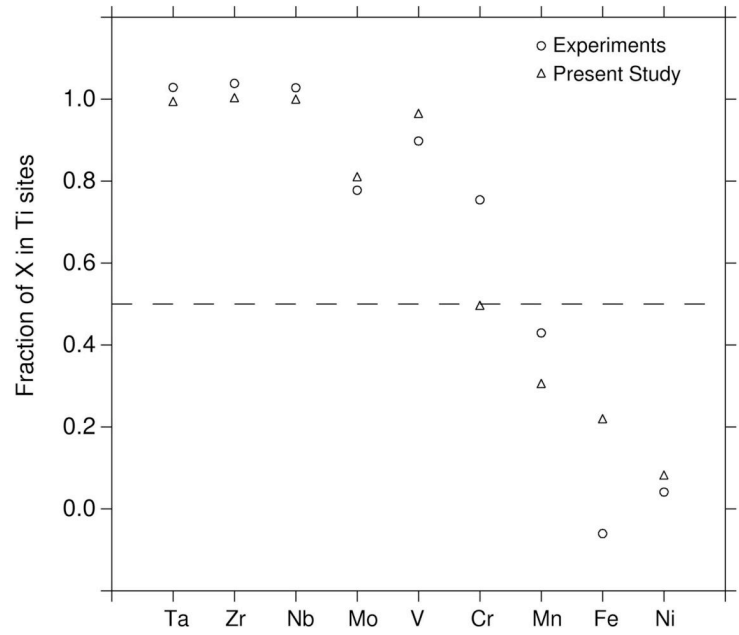

Fig. 4. Comparison between calculated and experimentally measured [27] site occupancy of X={Ta, Zr, Nb, Mo, V, Cr, Mn, Fe, Ni} in $L1_0$ $Ti_{0.47}Al_{0.51}X_{0.02}$ alloys at 1173K.